# Bright-field Nanoscopy: Visualizing Nano-structures with Localized Optical Contrast Using a Conventional Microscope

*Swathi Suran, Krishna Bharadwaj, Srinivasan Raghavan and Manoj M. Varma\**

Ms. S. Suran, Mr. K. Bharadwaj, Prof. S. Raghavan, Prof. M. M.Varma
Centre for Nano Science and Engineering, Indian Institute of Science, C.V. Raman Avenue, Bangalore 560012, India.
E-mail: mvarma@cense.iisc.ernet.in

**Abstract**

Most methods for optical visualization beyond the diffraction limit rely on fluorescence emission by molecular tags. Here, we report a method for visualization of nanostructures down to a few nanometers using a conventional bright-field microscope without requiring additional molecular tags such as fluorophores. The technique, Bright-field Nanoscopy, is based on the strong thickness dependent color of ultra-thin germanium on an optically thick gold film. We demonstrate the visualization of grain boundaries in chemical vapour deposited single layer graphene and the detection of single 40 nm Ag nanoparticles. We estimate a size detection limit of about 2 nm using this technique. In addition to visualizing nano-structures, this technique can be used to probe fluid phenomena at the nanoscale, such as transport through 2D membranes. We estimated the water transport rate through a 1 nm thick polymer film using this technique, as an illustration. Further, the technique can also be extended to study the transport of specific ions in the solution. It is anticipated that this technique will find use in applications ranging from single-nanoparticles resolved sensing to studying nanoscale fluid-solid interface phenomena.

**Introduction**

Visualization of structures below the optical diffraction limit requires special super-resolution techniques which are now well developed in the case of fluorescence imaging. Such techniques include Stimulated Emission Depletion (STED) [1], Stochastic Optical Reconstruction Microscopy (STORM) [2] and Photo-activated Localization Microscopy (PALM) [3]. In many situations one may only need to detect the presence of a nanoscale feature without needing to resolve it. For instance, the ability to count single nanoparticles may be of use in studying stochastic protein-protein interactions at the few-molecule limit [4]. We use the term visualization to refer to the detection of spatial features without necessarily being able to optically resolve the detailed geometry of the feature. Among the far-field, non-fluorescent techniques used for visualizing nano-scale features, dark-field (DF) microscopy can detect single metal nanoparticles down to 40 nm based on the Localized Surface Plasmon Resonance (LSPR) effect [5].However DF microscopy requires much larger sizes with dielectric particles. Photothermal imaging (PHI) [6] can detect objects down to 1.4 nm but requires intense pump sources [7]. Spatial Modulation Spectroscopy (SMS) [8] can visualize nanoscale objects by using phase locked detection of the intensity modulation caused by the motion of the nanoscale object moving in and out of the focal volume. Detection of 50 nm metal particles was demonstrated using SMS and the limit of detection was projected to be around 5 nm. Recently, it was shown that scattering from single protein molecules can be obtained under optimal signal-to-noise conditions achieved through control of the illumination and detection parameters [9]. Here, we present a technique to visualize nanoscale surface features based on the optical contrast generated by the differential etching of a germanium (Ge) thin film (~30 nm thick) deposited on gold in the presence of water. Such ultra-thin Ge films on gold produce a strong thickness dependent color response [10]. These Ge films also get etched in water due to the dissolution of its oxide $GeO_2$ in water. Any nanoscale structure placed on the Ge film, for instance nanoparticles dispersed on the film surface, impedes the transport of water to the Ge film resulting in a differential transverse etch rate and consequently a local color difference. The lateral etching of the underlying Ge film further amplifies the dimensions of the feature/structure allowing the local contrast to be observed using a conventional microscope with bright-field illumination [See Fig. 3]. This technique works for applications ranging from visualisation of single nanoparticles to extended nanostructures such as graphene grain boundaries and defect lines as thin as 2-5 nm. There has been significant recent interest in using graphene and other 2D materials as desalination [11] and filtration membranes [12, 13, and 14].Our technique can provide high resolution maps of water transport through such nano-membranes. This is possible by measuring the optical contrast between the nano-membrane and the background as a function of time, discussed in detail in further sections. The rate of change of optical contrast provides a measure of local water transport rates across the nano-membrane. We demonstrate this ability by measuring the volume rate of transport of water across a 1 nm thick polymer film deposited on top of the Ge layer. The technique can also be extended to probing the transport of ions in an aqueous solution. Also, anisotropic etch rates have been reported for metal nanoparticles deposited on crystalline Ge (100 orientation) in the presence of water [15]. The metal nanoparticles catalyse the oxidation of Ge to $GeO_2$ in their neighbourhood leading to local enhancement of etch rates. We have observed a similar enhancement of lateral etch rates in our films in the presence of metal (Ag) nanoparticles (40 nm dia.). Based on this data, we estimate that this technique should enable the visualization of single nanoparticles down to about 2 nm.

**Mechanism of visualization**

Anti-reflective (AR) coatings are necessary in several applications where losses from reflection of incident optical radiation need to be eliminated[16].Reflective losses from metallic surfaces can be eliminated by depositing ultra-thin films of absorbing dielectrics on the metal surface [17, 18, 19]. Mikhail et al. [10] used ultra-thin Ge films (5-25 nm thick) deposited on gold to achieve anti-reflection condition for enhancing solar energy conversion efficiency. Germanium readily forms an oxide $GeO_2$, which dissolves in water [15] leading to progressive reduction of the Ge film thickness upon immersion in water. Concomitant with the reduction of Ge film thickness, the AR condition shifts leading to a change in color of the film, as described in [10].Our device consists of a 30 nm thick Ge film deposited on an optically thick gold film on a Silicon substrate. Any micro- or nano-scale structureon such a device produces a local difference in the water transport rate which translates to a local color difference due to differential transverse etch rates, [See Fig. S4 in Supplementary Information (SI) text, Section 5]. These thickness differences are significant enough to produce local color contrast differences due to thin film interference. A simultaneous lateral etching of the Ge film amplifies the lateral dimensions of the nanoscale feature making the color contrast observable under a regular bright-field microscope without any external modifications. This optical contrast can be enhanced using Differential Interference Contrast (DIC) mode as seen in Figure 3.

**Results**

We fabricated devices with Ge thicknesses ranging from about 5 nm to 30 nm to extensively characterize their color response and etch rates in water. The details of this characterization effort are provided in the Methods section and SI text, sections 1, 2 & 4.Fig. 1(b) shows the experimentally measured shift in the reflectance spectra of one such device as the Ge film progressively got etched in water. Bright-field images as well as reflectance spectra were simultaneously acquired using a 63x water immersion objective. The experimental data matched well with the expected spectra (Fig. 1 (c)) computed using a transfer-matrix method [SI text, section 3].The experimental and computed reflectance spectra were respectively normalized to a maximum value of one. The shift of the reflectance spectra manifests as a change in color of the device (Fig. 1 (d)). It is also seen from Fig. 1 (d) that the position of the reflectance minimum red-shifts nearly linearly with increasing Ge film thickness. The data for our devices are consistent with previous reports [10]. It is evident that a thickness change of a few nm can produce appreciable color change.

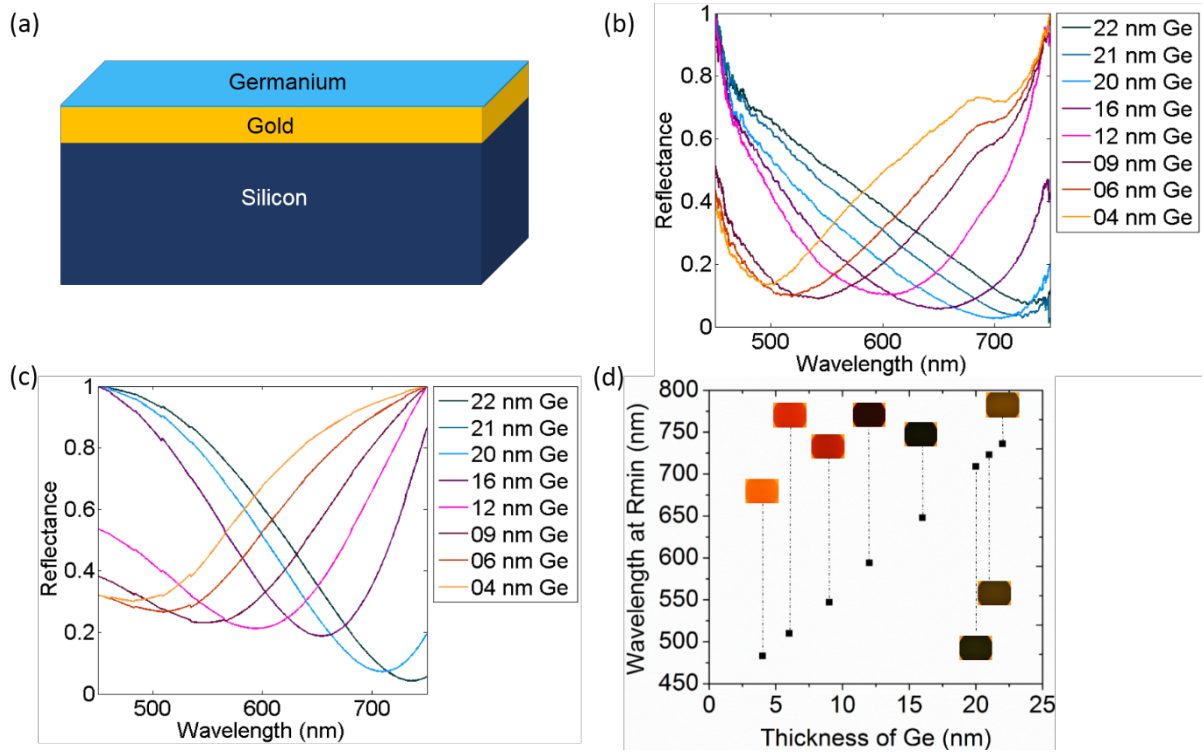

**Figure 1: Experimental and calculated reflectance spectra of the device. a.** Schematic of the device. **b.** The normalized experimental reflectance spectra for different thicknesses of Ge on Au **c.** The calculated reflectance spectra for the same thicknesses of Ge on Au. **d.** Corresponding reflectance minimum plotted for different thicknesses of Ge. The Bright-fieldmicroscope images as a function of Ge thickness are inserted next to each plotted data point. All images are captured at 50x magnification.

We investigated if defects and grain boundaries in CVD grown single layer graphene (CVD-SLG) can be visualized using this technique. While large morphological features like tears and folds of graphene are readily visible under a simple optical microscope, visualising structures such as grain boundaries and pinhole defects in graphene requires High Resolution Transmission Electron Microscope(HR-TEM) [20]and Scanning Transmission Electron Microscope (STEM) [21, 22] .We were motivated by the expectation that Ge film directly underneath a defect may etch at a faster rate due to possibly higher rate of water transport through the defects relative to the non-defective area. The resultant color contrast would allow us to conveniently identify graphene grain boundaries (GGBs) and other defects using a regular optical microscope instead of an HR-TEM or STEM. In order to test this hypothesis, a large area graphene film was initially synthesised on a Copper foil in a CVD chamber and subsequently transferred onto our device. The details of the growth conditions and transfer procedure are described in the methods section. Raman measurements confirmed the presence of single layer graphene on the device [SI text, section 7].

About 20 µL of deionised water (DI water) was drop-casted on the transferred graphene film, covering it entirely. We then used a 63x water immersion objective to observe the time-course of water transport through these graphene films. As time progressed, an etching pattern, shown in Fig. 2(a), representing a grain size of about 50 µm became evident at first in the Ge film. The average grain size of our graphene monolayer is of the order of 2-5 µm[SI text, section 7] whereas the grain size of the Cu foils on which the monolayers are grown have a grain size of 50 µm. Therefore, the 50 µm scale features observable initially are presumably due to the transfer of Cu grain boundary pattern onto the graphene monolayer which is grown on top. These high energy boundaries on the Cu surface are responsible for many more nucleation sites for graphene grains and therefore a large density of graphene grain boundaries per unit area is expected around them [23]. The graphene monolayer is thus expected to be more defective in these regions. Our data thus reveals two aspects about the graphene monolayers. One, the grain boundaries in the Cu translate to defects in the graphene monolayer and two, these boundaries are more permeable to water. After approximately 2hrs of etching, the GGBs (2-5 nm in width) were clearly visible (Fig. 2 (b) and (d)) with a noticeable contrast due to transverse and lateral etching of the underlying Ge as discussed in the previous sections. The tears in the film, folds in Graphene along the Cu rolling marks etc. can also be clearly distinguished (Fig 2(c)). The color balance of figures 2(b) and (c) have been adjusted for greater visual contrast using open source software ImageJ [24]. The raw images of 2 (b) and (c) are shown in the SI text, section 8. The complete time course of the etching process is presented as supplementary video 1. The differential etching observed here is likely to be a combination of enhanced oxidation [25] and subsequent dissolution of the Ge film underneath defective areas of graphene.

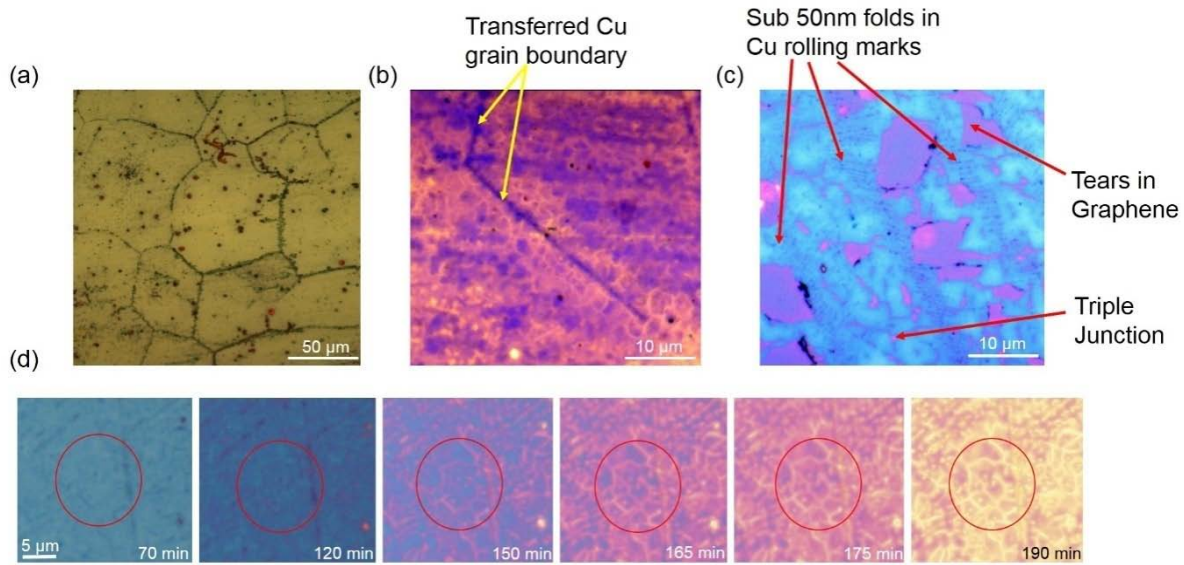

**Figure 2: Observation of defects in graphene. a,** A bright-field image showing defects in graphene along the Cu grain boundaries. **b.** Observation of grain boundaries in Graphene along the Cu grain boundary (indicated with yellow arrows). This Bright-field (BF) image has been post processed to improve visualization. The boundaries encircling grains (GGBs) appear yellow in the image. We clearly see that the typical grain sizes are about 2-5 µms.**c.** A post processed BF image where GGBs, tears in graphene, triple junction and folds along the Cu rolling marks are clearly visible. The Ge beneath the GBs/ other defects etch faster (pinkish region) than at the grains themselves (large blue regions). **d,** Shows the evolution of the appearance of GGBs with time. The raw images of b. and c. are shown in the SI text, section 8].

GGBs allow lateral etching of the underlying Ge thin film, amplifies the true widths (2-5 nm) to trenches of 20 nm in width as measure in AFM, shown in Fig.3. The ability to detect features of dimensions about 2-5 nm using an optical microscope, points to the potential of this technique for nanoscopic applications. Raman measurements performed after the etching process confirmed the presence of an intact single layer graphene film post-water etch [SI text, section 7]. In addition to features on single layer graphene, we were also able to visualize other nanoscale features such as photoresist lines [Data not shown] thus demonstrating the general applicability of this technique.

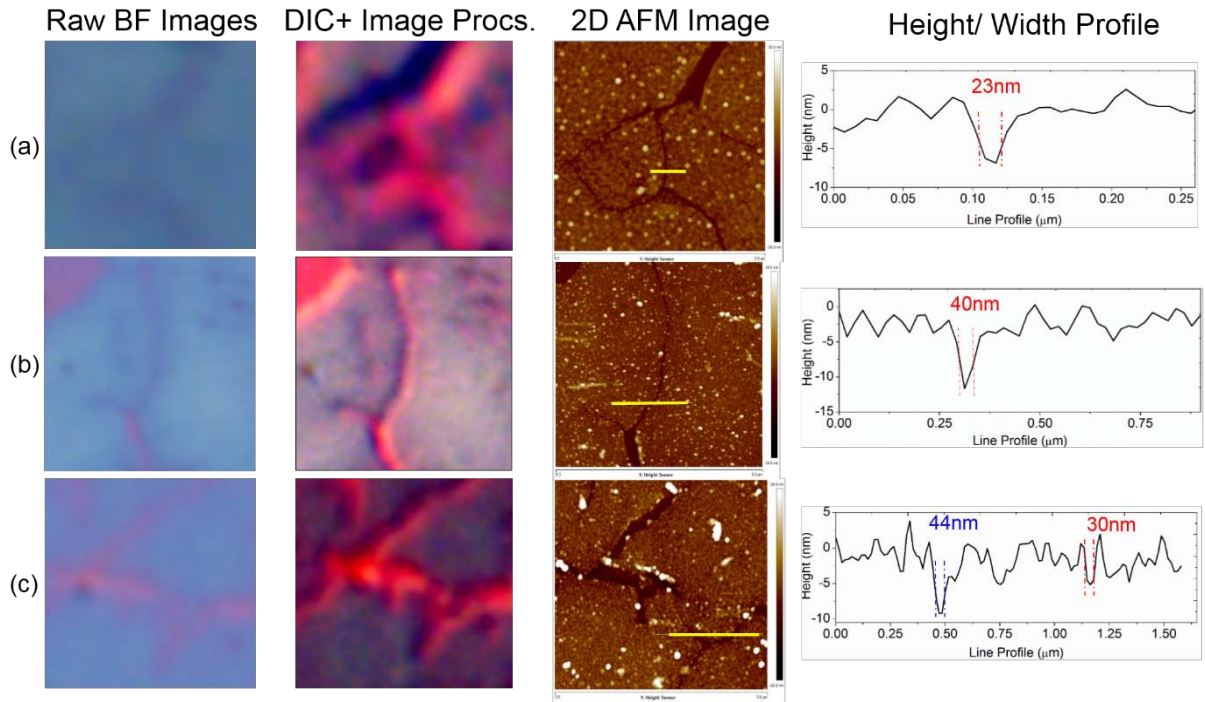

**Figure 3: Imaging grain boundaries (GB) from different SLGs. a.** GBs visualized in SLG grown with low defect densities. First column shows raw image obtained from Bright-field (BF) mode in an optical microscope, the second column contains the color balance adjusted version of the images obtained using the DIC mode. The third and the fourth column shows the corresponding AFM data, a 2D AFM scan and the height profile measuring the width of the etched Ge trench across the solid yellow line marked in the AFM image. **b.** and **c.** Show visualization of GBs in SLGs grown with higher defect densities in BF and DIC mode with the AFM data supporting it. The optical images are cropped and magnified from a 100x image taken in both BF and DIC modes. The raw BF and DIC images are shown in the SI text, section 8.

In order to investigate the use of this technique to measure water transport through ultra-thin (1-5 nm thick) membranes (referred to as nano-membranes in this article), we fabricated a microarray pattern of a single layer of a polyelectrolyte [26], PAH (Poly(allylamine Hydrochloride)) on top of our device as shown in Fig. 4(a). The thickness of a single layer of PAH is about 1 nm [SI text, section 9]. The time course of water etching of this device with the PAH pattern was measured using the 63x water immersion objective. Initially, the polymer microarray was not visible against the background (Ge device) due to low optical contrast as seen in Fig. 4 (b). However due to the difference in the rate of water transport through the polymer nano-membrane relative to the background (bare Ge) the optical contrast gradually increases due to the increasing difference in the underlying Ge film thickness. The complete sequence of etching is provided as supplementary video 2. The rate of change of color (or more precisely, the spectral position of the reflectance minimum) of the region containing the polymer, relative to the background is a measure of the transport rate of water through the polymer. For e.g., if water can pass unimpeded through the membrane, we would expect the rate of color change between the polymer and the background to be the same. On the other hand, if the polymer offers high resistance to the transport of water, we would expect no change in the color of the region containing the polymer while there would be a gradual change in the background color until it reaches the gold film which acts as an etch stop. In principle, measurement of color differences can enable the extraction of transport rates of water in these systems. In practice, precise quantification of transport rates using color values is challenging due to a) non-linear behaviour of color values with respect to spectral shifts and b) a detailed quantitative model to relate Ge etch rate differences between the region under the membrane and the background. It is possible that different membranes in addition to offering varying resistance to water transport also change the permeability of oxygen which would affect the $GeO_2$ content under the membrane. It is $GeO_2$ which dissolves in water and not Ge directly. A precise quantitative estimation must take care of these effects rigorously. Although at present, we do not have such a rigorous model, a naive model [SI text, section 9] yields a transport rate of 510 pL/min through the 1 nm thick PAH layer. Using appropriate quantitative models, this technique may be useful for fundamental studies of water transport through nano-scale membranes [12, 13, 14] as well as high resolution mapping of water transport heterogeneity in such membranes. As described later in the article, this technique could also be extended to measure the transport of ions through nano-membranes using selective intermediate polymer layers between the nano-membrane and the Ge film.

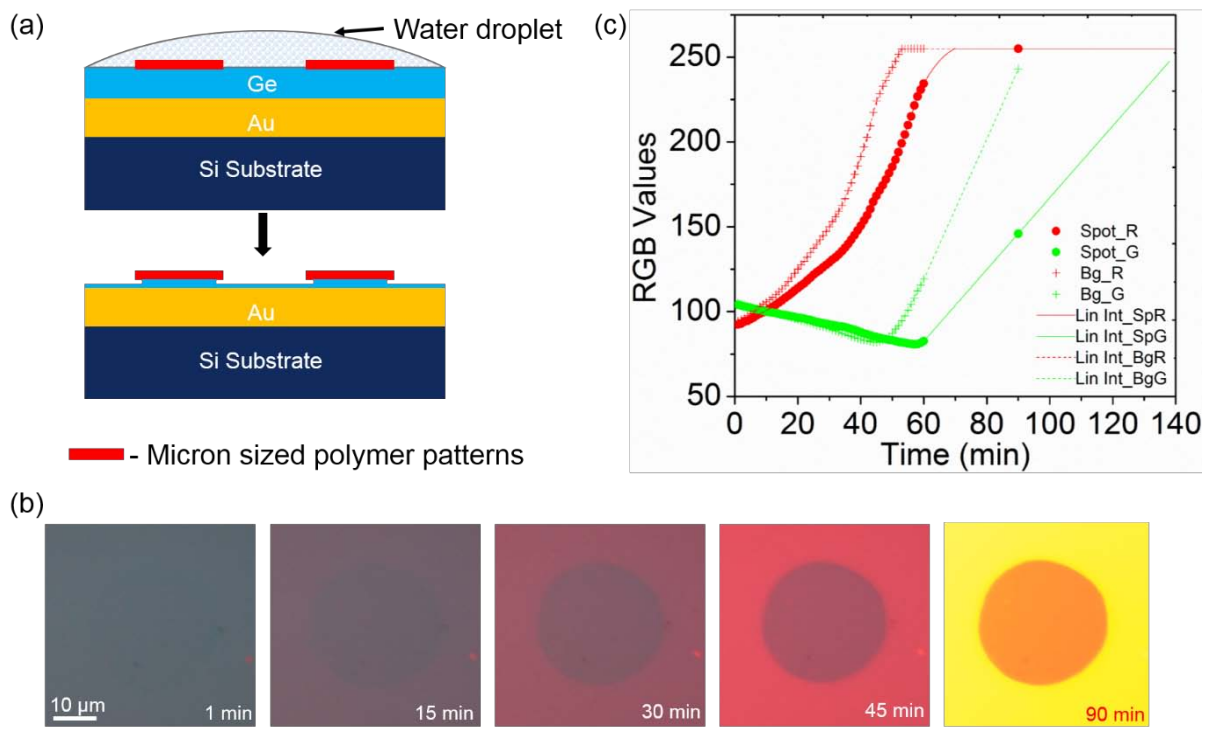

**Figure 4: Differential water transport through nanometer thick polymer films. a.** Schematic showing polymer films patterned (spot) on the Ge device undergoing etching in the presence of a water droplet. After 90 mins of etch, Ge at the background has almost completely eroded when compared to the Ge beneath the polymer. **b.** Sequence of images showing as Ge etched at the polymer pattern (spot) and the background (Bg). All optical images have been captured using a 63x water immersion objective. The time stamp is marked in each of these images indicating when the image was taken with respect to the first image. **c.** A plot of R, G and B obtained over time from the spot and background (bare Ge) is compared to that of the bare Au deposited.

As a final demonstration of the utility of this technique in nanoscopy, we imaged isolated metal nanoparticles using this technique. To ensure adhesion of the particle to the Ge film during the rather long etching process (~ 90 minutes), we used triangular and disk shaped Ag nanoparticles with size ~40 nm [SI text, section 10]. We hypothesized that these NPs with larger surface area of contact would adhere to the Ge layer without the use of additional functionalization layers. It has been shown earlier that metal NPs can lead to enhanced etch rate of Ge (100 orientation) crystal planes due to the catalytic action of the metal NPs. This phenomenon led to a lateral expansion of the feature size as revealed by electron microscopy [15]. Our work deals with Ge thin films as opposed to crystalline Ge wafers. Nevertheless, we observed similar enhancements of lateral feature size of 40 nm Ag

NPs deposited on our device after water etching for about 90 minutes. Correlated optical and AFM images in Fig. 5, demonstrate the visualization of 40 nm Ag nanoparticles using our technique. Lateral etching resulted in a spatial magnification factor of about 10 for 90 minutes of etching. This lateral magnification factor implies that a 4-5 nm particle would produce a feature size of about 40-50 nm after etching which should be readily visible in the microscope. We tried to verify this limit experimentally using spherical Au NPs and CNTs. However, the optical contrast was not sufficiently high most likely due to the lack of adhesion of these particles to the non-functionalized surface. Poor adhesion would lead to motion of the particle during the etching process and consequent reduction (wash-out) of the differential optical contrast. The results of visualization of 2 nm dia CNT is shown in the SI text [SI text, section 11]. We believe that visualization of nanoparticles or objects down to 5 nm should be possible by a) using adhesion layers to reduce fluctuations of the position of the particle during the etch process and/or b) increase the etch rate of the Ge films by increasing the $GeO_2$ content during film deposition or post-deposition oxidation. Such studies are presently underway.

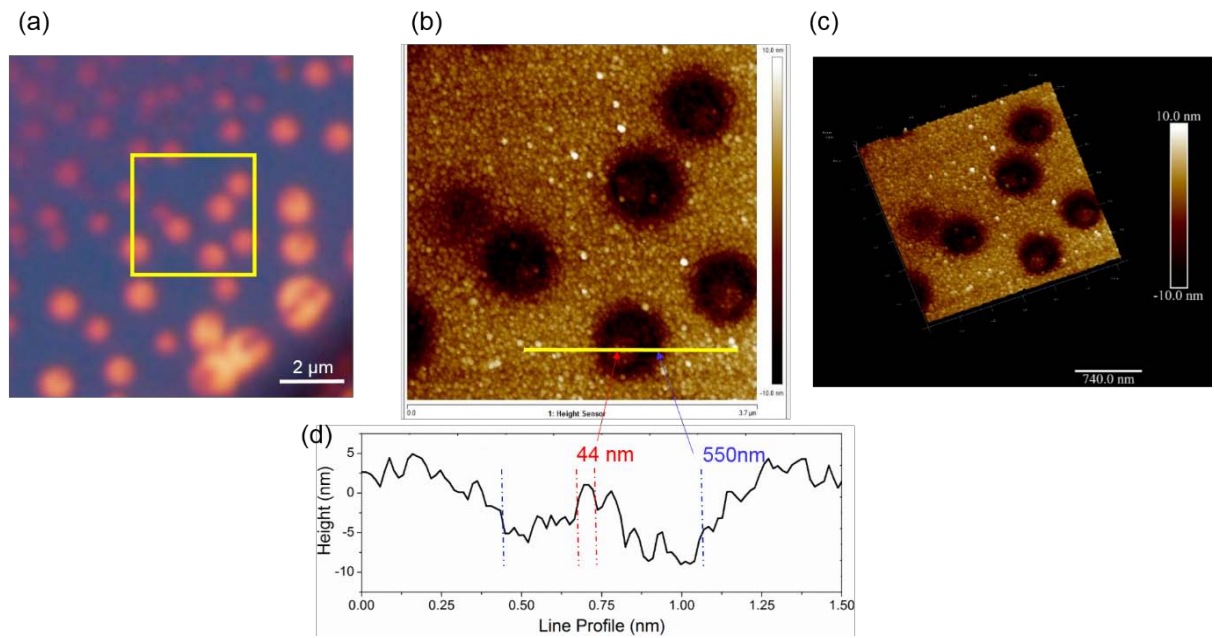

**Figure 5: Visualizing 40nm Ag NPs. a,** Optical image showing NPs which caused the substrate, Ge to etch laterally around the particle to form circular pits. **b,** Shows the AFM of the area in the yellow box in **a**. **c** and **d** are the 3D view and line profile along line marked in **b.**

## Discussion

The working principle of our technique is the combination of the facts that Ge thin films on gold can produce strong color response and that $GeO_2$ formed readily on the Ge film surface can be etched with water which is a benign solvent permitting imaging using water-immersion microscope objectives. The etch rate of $GeO_2$ has been reported to range from $10^{-2}$ nm/min to $10^4$ nm/min depending on the oxidative state of Ge [27].We have determined the etch rate in our system to be around 0.3 nm/min [SI text, section 8]. The $GeO_2$ content of the Ge films prepared in our RF sputtering tool was determined to be about 28% using XPS [SI text, section 6]. Depth resolved XPS confirmed that $GeO_2$ is present only near the surface and the bulk contains elemental Ge [SI text, Section 6]. This implies that the mechanism of differential etching involves both differential oxidation rates as well as differential water transport. The relative significance of these two effects must be clearly understood to use this technique for quantitative transport studies. However, the technique can be used as a nanoscopic visualization tool even in the absence of such a quantitative model. This technique also probes the heterogeneity in the water transport through nano-membranes which will be useful in the design of membranes for applications such as water desalination. One obvious limitation of this technique is that it is currently restricted to probe transport of water alone. However, by using polymer thin films which selectively etch in the presence of specific ions, it would be possible to measure the transport of these ions. For example, the solubility of $PAA-Ca^{2+}$ increases with increasing $Na^+$ concentration in water [28].By depositing an intermediate layer or $PAA-Ca^{2+}$ sandwiched between the nano-membrane and the Ge film, one can use this technique to probe the transport of Na+ ions through the nanomembrane as described in the SI text section 12. The estimated limit of detection of this technique for visualization of nanoparticles is about 5 nm which should permit single-molecule resolved studies by using functionalized NPs or QDs (4-5 nm). At this limit, it will also be possible to observe single macromolecules or polymers optically, without the use of fluorescent labels. In conclusion, we have demonstrated a nanoscopic visualization technique with significant potential in areas ranging from fundamental studies of solute transport in nanoscale membranes to single-molecule resolved sensing of molecular interactions.

## Methods

**Device Fabrication:** A Si wafer was cleaned in Piranha solution (3:1 of Sulphuric acid and Hydrogen Peroxide) and used as the substrate. All thin films weredeposited in a Tecport sputtering unit. A 150nm, optically thick Au film was sputtered on the Si substratewith a

10nm thin layer of Cr as an adhesion promoter, followed by which Ge films of thicknessrangingfrom 5-30 nm were deposited on theAu film. Au and Ge were deposited at the rate of about 5Å/sand 2Å/sec respectively under a vacuum of $10^{-6}$ Torr. A bare Si wafer was used in addition to the metal coated substrate for ellipsometric determination of deposited Ge film thickness. In some cases the Ge film was also patterned permitting the measurement of Ge film thickness using an AFM.

**Measuring reflectance spectra from the microscope:** The reflectance spectra of the devices were measured using an Ocean Optics modular fibre optic spectrometermodel USB 2000. The receiving optical fibre was secured to the eye-piece of an upright microscope (Model Olympus BX51M) to collect the reflected light from the device. The reflected light collected by the optical fibre was sent to the spectrometer which was controlled by the Ocean Optics software via a USB interface. The calibrated reflectance spectra were obtained using the Ocean Optics software and the corresponding color images were simultaneously captured by a CCD (Model: Olympus DP73) attached to the camera port of the microscope.

**Synthesis and transfer of Graphene.** Large area Graphene was synthesised on Cu foil in a CVD chamber with precursor gas Methane in presence of Hydrogen. All the graphene films were grown at a temperature of 1000 °C under a total pressure of 4 Torr. Graphene grown on the Cu foil was transferred on to Ge thin film devices using the conventional wet transfer process. The graphene-Cu composite was initially coated with a thin PMMA support layer followed by a soft bake at 180 C. The underlying Cu was subsequently etched using ammonium persulphate solution. The free standing graphene-PMMA composite was then fished out using an oxidized Si wafer. After rinsing with de-ionised water for a few times, the graphene layer was finally transferred on to the Ge thin film devices. The final graphene-on-Ge film was then vacuum dried using a desiccator for 3 hours to avoid trapped water between graphene and the Ge thin film surface.

Further details of the experimental protocols and characterization procedures are provided in the supplementary document, SI text sections 1-7.

# SUPPORTING INFORMATION

**Bright-field Nanoscopy: Visualizing Nano-structures with Localized Optical Contrast Using a Conventional Microscope**

*Swathi Suran, Krishna Bharadwaj, Srinivasan Raghavan and Manoj M. Varma\**

Ms. S. Suran, Mr. K. Bharadwaj, Prof. S. Raghavan, Prof. M. M.Varma
Centre for Nano Science and Engineering, Indian Institute of Science, C.V. Raman Avenue, Bangalore 560012, India.
E-mail: mvarma@cense.iisc.ernet.in

### Section 1: Fabrication and characterization of the device

A p-type <100> oriented silicon wafer was cleaned in piranha solution consisting of 3 parts of sulphuric acid and 1 part of hydrogen peroxide. The wafer was bathed in buffered hydrofluoric acid (BHF) to remove any native oxide present on the surface. Subsequently, chromium and gold films of thicknesses 10 nm and 150 nm respectively were deposited in a Tecport sputter coater. Chromium was sputtered to improve the adhesion of the gold to the Si substrate. No substrate heating was used during sputtering. Au was deposited at a rate of 5Å/sec under high vacuum of $10^{-6}$ Torr using DC power. Followed by Au, germanium was sputter deposited at a rate of 2Å/sec using RF power to attain different thickness varying from 5-30nm by changing the deposition time. For thickness optimisation, circular patterns of 100 and 200 µm were patterned by photolithography in EVG 620 double sided mask aligner using a dark field mask. Subsequently, germanium of different thicknesses was sputtered on these lithographically patterned substrates after which the photoresist was stripped off leaving Ge spots of 100 and 200µm diameter. The thickness of these patterns was measured in a Bruker Scan Asyst Atomic Force Microscope in the tapping mode. Parallally, Ge films were also deposited under exactly similar sputtering conditions on a bare Si wafer so that the thickness of Ge could also be measured optically using a Variable Angle Spectroscopic Ellipsometer (VASE) J. A. Woollam Co. M2000U. From Figure S1., it is seen that the thickness of Ge measured using Ellipsometry and AFM are in reasonable agreement.

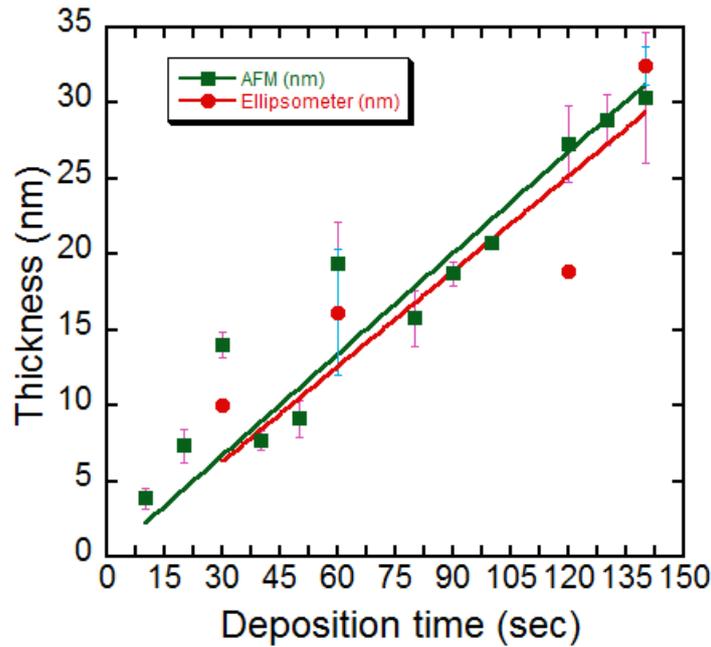

**Figure S1.** Correlation between Ellipsometer and AFM measurements of the thickness of Ge film deposited for different deposition times.

**Section 2: Reflectance spectrum measurement**

Different thicknesses of Ge on Au, give rise to an array of uniquely separated colors which were captured using an upright microscope from Olympus, BX51M in the bright-field mode. The color response is due to interference of the reflected light from the top and bottom interfaces of the Ge film. In order to measure the reflectance spectra, we used a table-top, miniature fibre-optic spectrometer by Ocean Optics, the USB2000 model. We mounted the receiving optical fibre of the spectrometer on the optical microscope such that it collect all the reflected light from sample through the eye-piece. The light from the eye-piece was carefully focused to the inlet of the optical fibre. The spectrometer was connected to the microscope's computer via the USB port. This set up enabled us to capture both the reflectance spectra and bright field images of the device simultaneously. The bright field images of the different thicknesses of Ge deposited on Au are shown in Fig 1(d) in the main text. All images were captured in the illumination from the Halogen lamp without any external filters. With increasing thickness of Ge, we observe a red-shift of the reflectance spectra, as expected [1].

**Section 3: Theoretical calculation of reflectance spectra**

We used a transfer matrix method to calculate the reflectance spectrum of the Ge thin film structure shown in Fig. 1 (c) of the main text [2]. The calculations were done using the MATLAB software package. The refractive index for Si and Au used in the simulation were obtained from an online RI data base [3], whereas the optical constants used for Ge were extracted from the ellipsometer data for a 30 nm thick sputtered Ge film. As seen from Fig. 1 (b) and (c) of the main text, the calculated spectra are in good qualitative agreement with the experimentally measured reflectance spectra.

**Section 4: Estimating etch rates of sputtered germanium thin films in water**

Ge film readily oxidises itself to become $GeO_2$ which is soluble in water [4]. The oxidation and consequently the oxide film is confined to the surface. All our experiments were carried out within 2-3 days from the time of deposition. Assuming that these films etch isotropically with water, we estimated the etch rates and discuss its association with the reflectance spectra in this section. Five different thicknesses of Ge, namely, 14, 20, 22, 26 and 30nm, were deposited on Au substrates in the array pattern described in section 1. The initial thicknesses were measured in an AFM. These samples were then etched in DI water for a duration of 15 mins and the thickness of the Ge layer was measured again using the AFM. The average etch rate observed for all these samples was about 0.3 nm /min.

A further study was performed where different devices, all with the same initial thickness, were etched for different durations. The height difference as a function of etch time was plotted as shown in Fig. S2. This study also yielded an average etch rate of about 0.3 nm/min. The reflectance spectra were captured as a function of time as the films underwent progressive etching in water. With the spectrometer mounted on the microscope, we were able to simultaneously record both the optical image as well as reflectance spectra. When recording the reflectance spectra from Ge patternsthe Field Stop (FS) of the microscope was adjusted such that only the light reflecting from the Ge region is collected by the spectrometer's optical fibre. The reflectance spectra as a function of etch time is shown in Fig. S3. It can be seen from Fig. S4 that the spectral shift due to Ge thickness change is about 13 nm/nm, i.e. 13 nm shift in wavelength per nm of change in Ge thickness. Given that even crude spectrometers can easily achieve nm level resolution of spectral features, this means that angstrom level changes in Ge film thickness due to etching can be quite easily measured.

Further, we note from Fig. S3, that the slope of the spectral shift curve (13 nm/nm) was fairly robust across several samples.

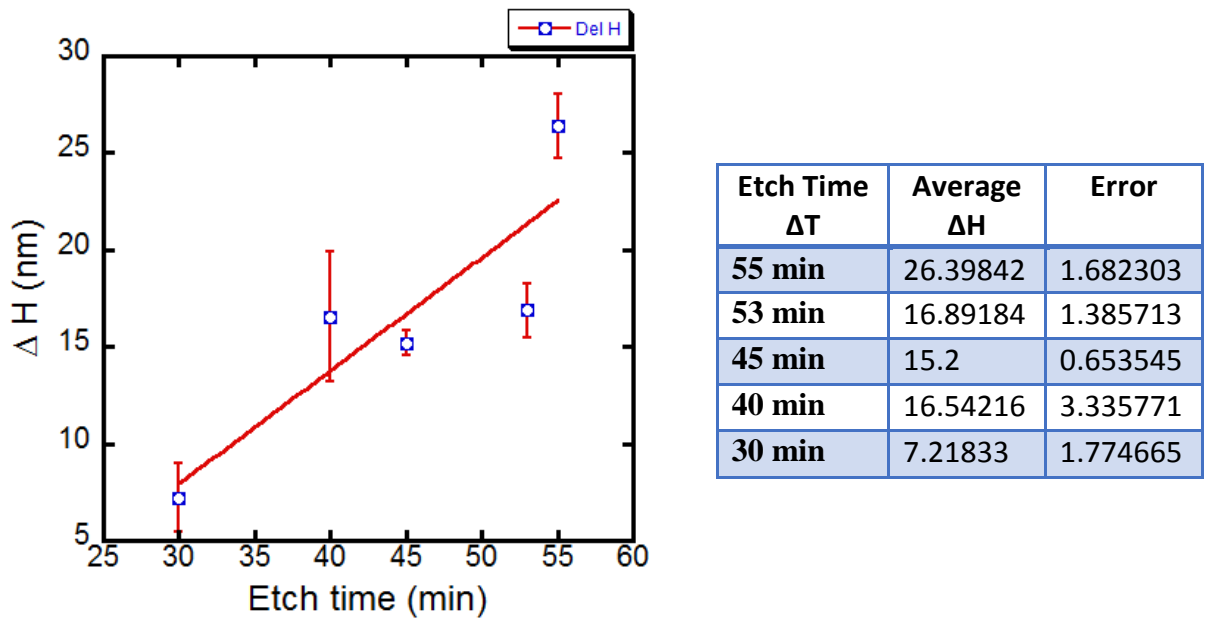

| Etch Time ΔT | Average ΔH | Error |
|---|---|---|
| 55 min | 26.39842 | 1.682303 |
| 53 min | 16.89184 | 1.385713 |
| 45 min | 15.2 | 0.653545 |
| 40 min | 16.54216 | 3.335771 |
| 30 min | 7.21833 | 1.774665 |

**Figure S2**: Change in Ge film thickness as a function of etch time.

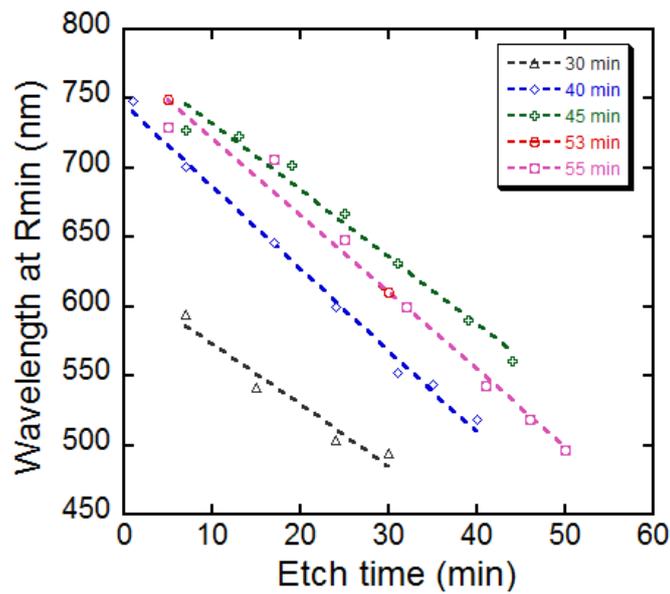

**Figure S3:** Spectral shift (shift of the position of minimum reflectance) of the Ge devices as a function of etch time in DI water

**Section 5: Mechanism of visualisation due to differential etch**

Ready oxidation of Germanium to $GeO_2$ and its dissolution in water and the capability of the device to pick up small changes in the Ge thickness [1] with a high contrast enabled the visualisation of nano-structures. Any nano-structure like nano-particles or graphene with defects on the device when underwent water etch, as shown in the Fig. S4 a. & b. manifested to local differences in etch of the underlying Ge which was easily visualized using a BF optical microscope with a large color contrast.

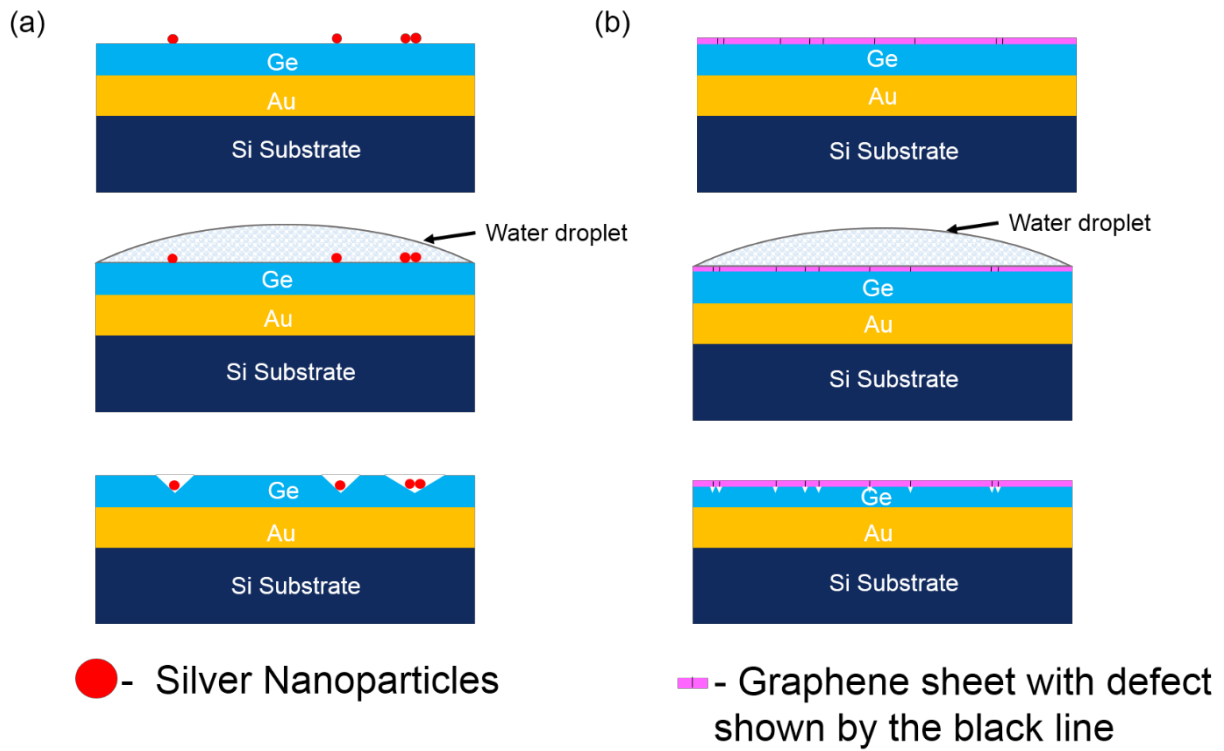

**Figure S4: a.** Ag NPs drop casted on the device undergoes etch in the presence of water, resulting in anisotropically etched pits in Ge. **b.** Graphene placed on the device. On water etching, the Ge beneath the GGBs (marked in black) etch faster when compared to the rest of the area (grains).

**Section 6: Analysis of native oxide on Ge using XPS**

As our technique strongly depends on the oxide content in the Ge film, we carried out studies using X-ray Photoelectron Spectroscopy (XPS, Model: Axis Ultra) to determine the composition of native oxide in the sputtered Ge films. Fig. S5 (a) shows the Ge 3d high resolution spectrum which indicates the presence of two oxidation states of Ge. The composition was estimated to be 49.2% Ge, 22.5% GeO and 28% $GeO_2$. The experimental plot in the Ge 3d spectrum comprises of two major peaks as seen in Fig. S5 (a). The large peak is associated to elemental Ge at 29.3 eV and deconvolution (peak-fitting) shows the presence of Ge in two different oxidation states, GeO (+2) at 30.9 eV and $GeO_2$(+4) at 32.5 eV. Further, depth-resolved XPS was carried out to determine the oxide present at the surface and how much it extends through the depth of the film. Depth resolved XPS revealed that the oxide (GeO$_2$) is confined to few nanometer thickness from the top surface of Ge. Fig S5 (b) shows Ge 2p $_{3/2}$ component of the Ge 2p high resolution spectrum. The Ge 2p before etch spectrum shows a significant presence of Oxygen as a second peak at 1220.4 eV other than the peak due to elemental Ge at 1217 eV. With consecutive etches, it was seen that the intensity of the peak at 1220.4 eV diminished as the Oxygen content decayed through the thickness of Ge. This implies that the mechanism of etching involves oxidation of Ge to GeO$_2$ due to dissolved O$_2$ content in the water followed by dissolution of GeO$_2$ due to water. Further studies are underway to control the GeO$_2$ content of the film across the depth to achieve etch rates of the order of 5 nm/min. The peak-fitting and other analysis of Ge thin film was carried out using the open-source software available online, CasaXPS [5].

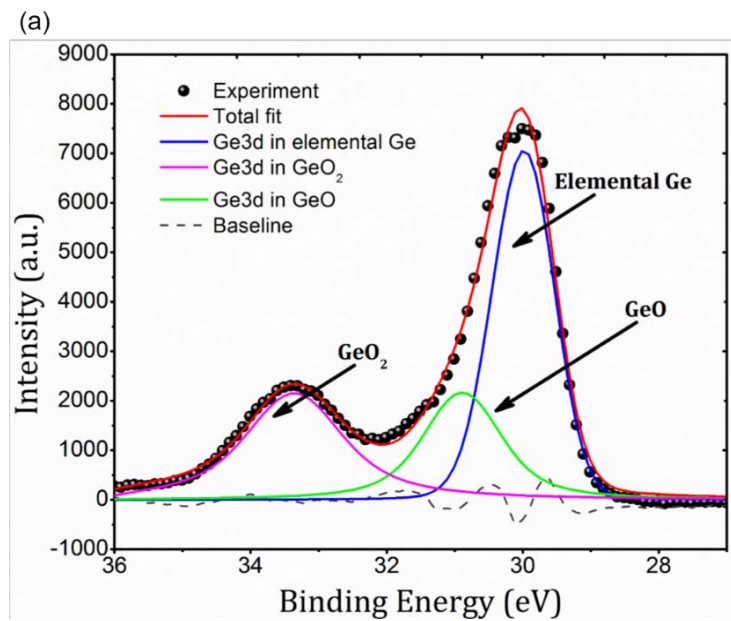

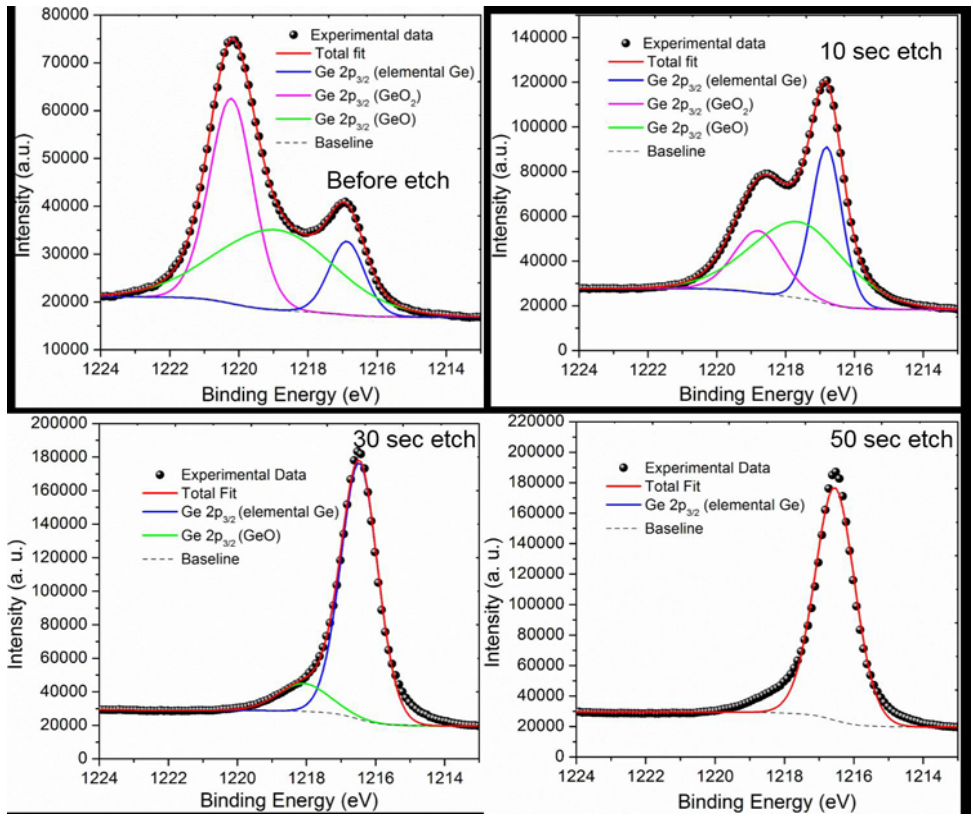

**Figure S5: a.** High resolution XPS spectrum of Ge in the $3d_{5/2}$ energy region showing the presence of native oxide on Ge. **b.** Depth resolved XPS showing the decay of oxygen content through the thickness of Ge. At the end of 30 sec etch cycle, the composition of the film was 100% elemental Ge.

**Section 7: Raman and SEM measurements of graphene layer on germanium**

Figure S6 shows the typical Raman spectrum from the single layer graphene layers transferred on to the Ge film. The Raman spectra were obtained using a 535 nm laser in a Horiba LabRAM HR. Three important peaks are noted in the figure as D (1350 cm$^{-1}$), G (1580 cm$^{-1}$) and the 2D (2750 cm$^{-1}$). While the 'D' intensity is quantitative measure of the defect density in graphene under the laser, the 2D peak width is an indicator of the number of layers. Low relative 'D' peak intensity ($I_D/I_G < 0.2$ and $I_{2D}/I_G > 3$) and a small 2D peak width (FWHM of 2D < 30 cm$^{-1}$) confirm the presence of good quality single layer graphene. Raman measurements were made before and after the DI water etching experiments to check if the etching caused any significant change in the graphene layer. Raman data shown in Fig. S7 confirms the presence of graphene after the etch experiments and additionally, no significant difference in the 'D' peak intensity ratio, $\Delta\ I_D/I_G < 0.1$ indicates that the graphene is not significantly modified by the etching process.

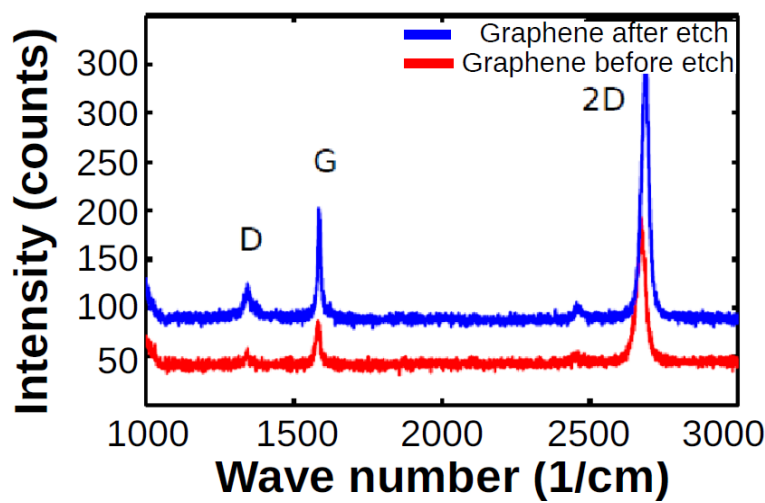

**Figure S6:** Raman spectra from single layer graphene transferred on the Ge thin film device. The spectra were obtained before and after the DI water etching of the underlying Ge film.

An estimate of the graphene grain size was obtained by analyzing the scanning electron microscopic (SEM) images of the grown grapheme just before coalescence of the film. Figure S7 shows a representative SEM image indicating that the graphene grains are mostly hexagonal with sides of length of 2-5 µm.

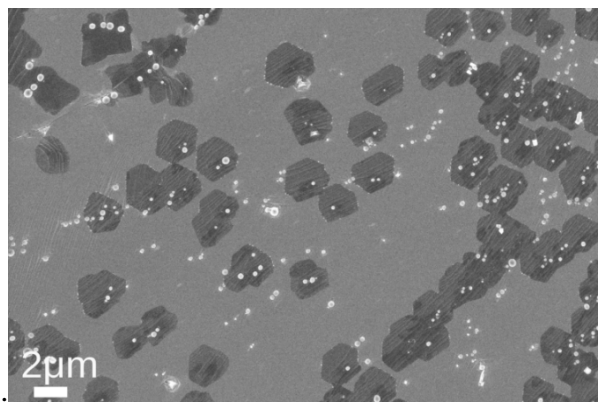

**Figure S7:** SEM image of Cu surface showing graphene grains before coalescence. Grains are mostly hexagonal shaped with sides of length 2-5 µm.

**Section 8: Original/ raw data of the images produced in the main text.**

Figure 2 in the main text illustrates the first observation of GGBs and also other anomalies in the transferred graphene. The images in the main text were obtained after changing the color balance of the raw image for optimal visual detection of the features. The original images before adjusting the color balance are provided in this section. The color balance of the raw images where adjusted using the Adjust – color balance option in the open source software package ImageJ. The raw and color balance adjusted images are shown side by side in Fig. S8 below.

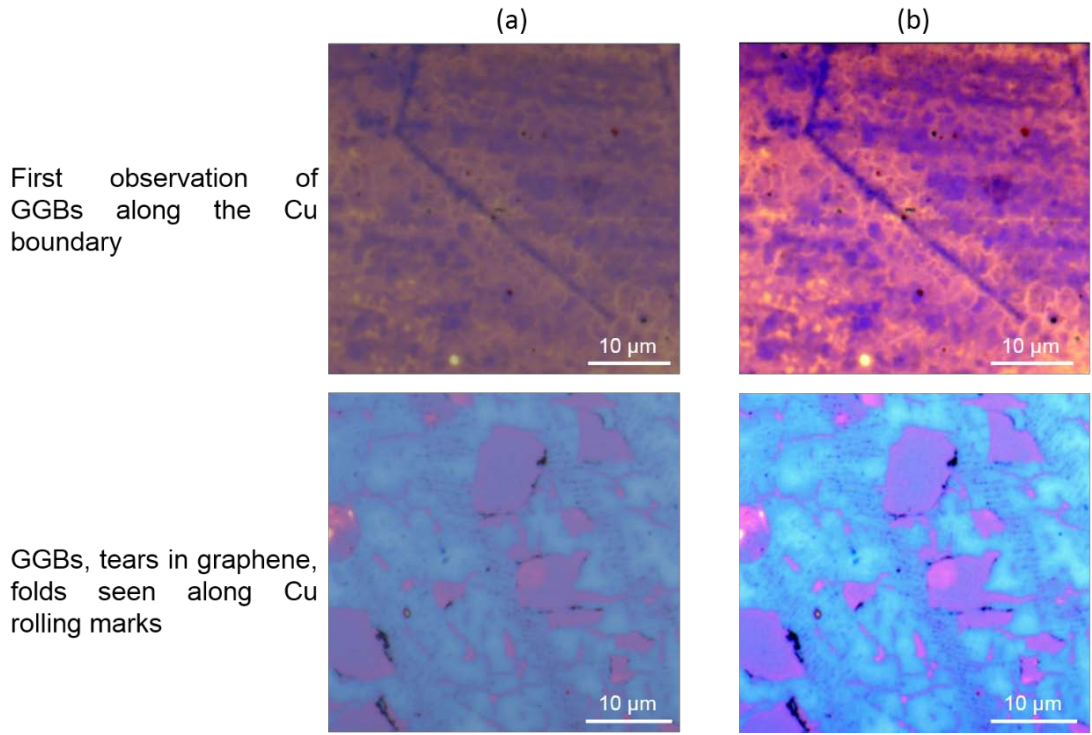

**Figure S8: Post processed raw images. a.** Raw and **b.** color balance adjusted images where grain boundaries and other defects are seen in the single layer graphene.

In Fig. S9 below, the encircled area in the time sequence image set in Fig 2 (d) from the main text has been magnified for better visualisation of the grain boundaries. Note that these images have not undergone any color balance adjustment. The grain boundaries have been marked with red solid lines to delineate them.

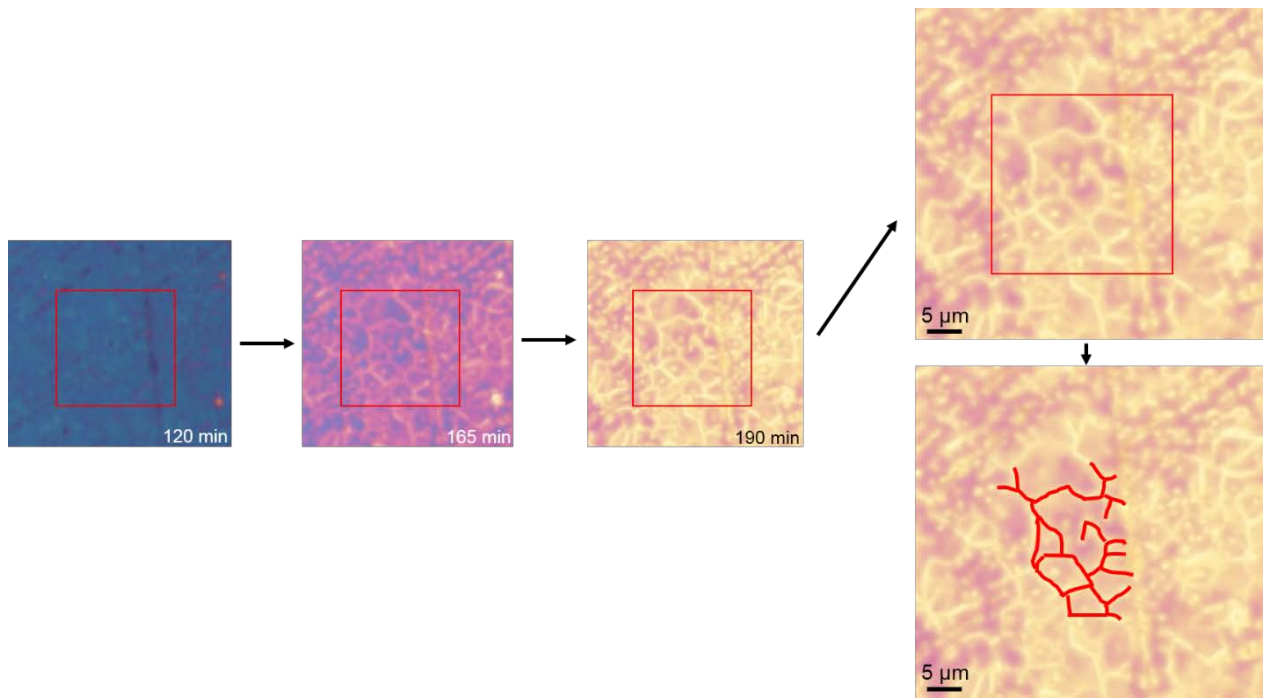

**Figure S9:** Magnified images of the region with graphene grain boundaries (Magnified from Fig. 2 (d) of the main text.

As mentioned in the main manuscript, Differential Interference Contrast (DIC) imaging mode provided better optical contrast compared to Bright-field imaging mode. Fig. 3 in the main text shows the raw BF images as well as DIC images after adjusting color balance. Fig. S10 below shows the raw DIC images and the comparison of raw BF and DIC imaging modes.

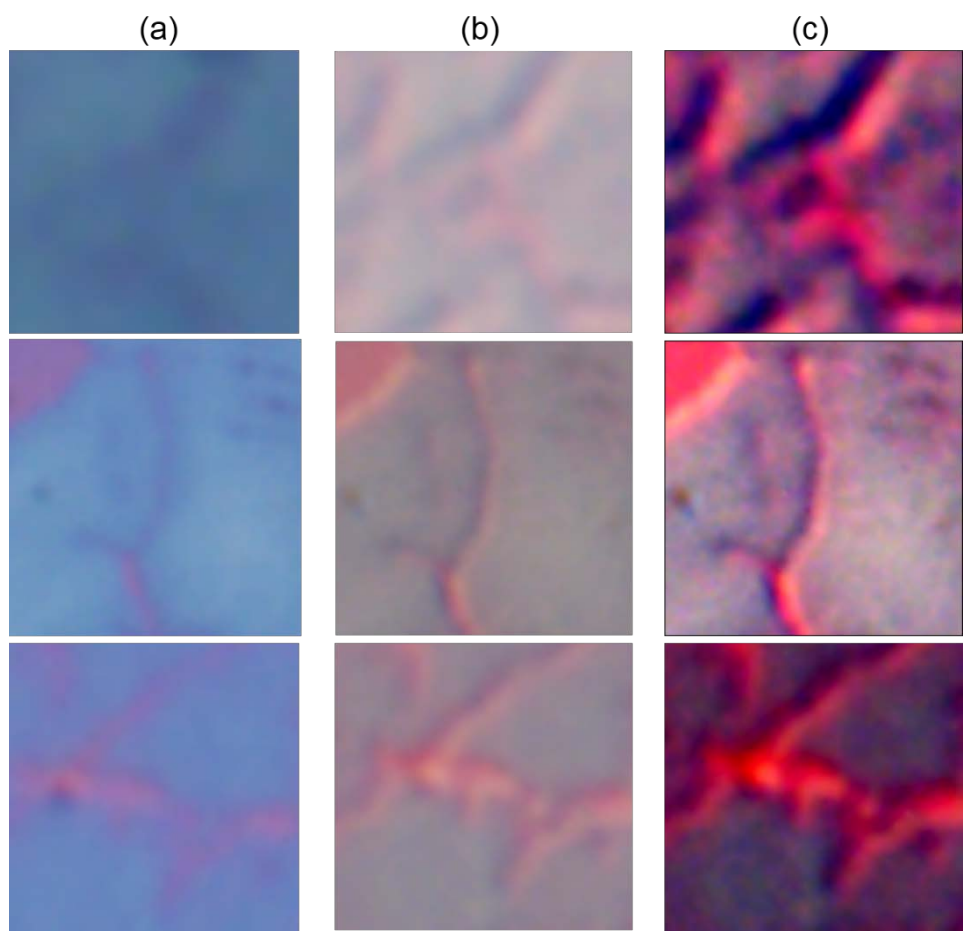

**Figure S10:** Comparison of raw BF and DIC imaging modes for nanoscopy. **a.** Raw BF images where GGBs from graphene grown under different growth conditions. **b.** Raw DIC images from the same location as in **a.b.**Processed (color-balance corrected) images showing improvedvisualisation of GGBs.

**SVideo 1:** Time course of water transport through GGBs

**Section 9: Differential water transport through ultra-thin polymer layer**

Polymer nano-membranes (single layer of PAH, ~1-2nm, as described in the main text) patterned on 30 nm Ge devices became visible with increasing optical contrast as the differential etching of the Ge film progressed. We measured the step height due to polymer to be around 1 nm before the beginning of the etch process. After 90 minutes of etch the step height increased to about 2.74 nm when measured in AFM as seen in Fig. S11. The increase in height is because of the Ge film directly underneath the polymer layer not etching as fast as the background, as the polymer layer impedes the access of the etchant (water) to the Ge film. From these height differences, we can estimate that the etch rate of background Ge (not under the polymer) is about 0.333 nm/min whereas etch rate of Ge through the polymer is 0.311 nm/min. The etching of Ge is approximately 6% slower through a single layer of polymer measuring about 1 nm in thickness. From this data, we estimated the volume flow rate of water through the polymer in the following manner. Let $\tau_p$ be the time it takes to etch one nm of Ge film directly under the polymer. Similarly, $\tau_b$ is the time for one nm of Ge film directly exposed to water to etch. Then we wrote $\tau_p$ as,

$$\tau_p = \tau_b + \tau_w$$

Where, $\tau_w$ is the time taken by water to permeate through the polymer layer. We estimated $\tau_p$ and $\tau_b$ from the inverse of the etch rates 0.31 nm/min and 0.33 nm/min respectively yielding $\tau_w$ to be about 0.19 minutes. This is the time it takes water to permeate a roughly 1 nm thkck polymer layer. Therefore the rate of water permeation is about 1 nm/0.19 minutes or about 5.1 nm/min. This is equivalent to a volume flow rate of 510 pL/min. It is quite obvious that this estimate is not a very precise one due to the uncertainties and variations in step heights between various polymer pillars. A precise quantitative estimate will require a rigorous model of color change or spectral shift as a function of water transport rate through membranes. Such a model should account for change in Oxygen permeability due to the nanomembrane as well as water transport through the membrane. However, in the present work we employed the crude model described above to illustrate the potential of technique in transport measurements through nano-membranes.

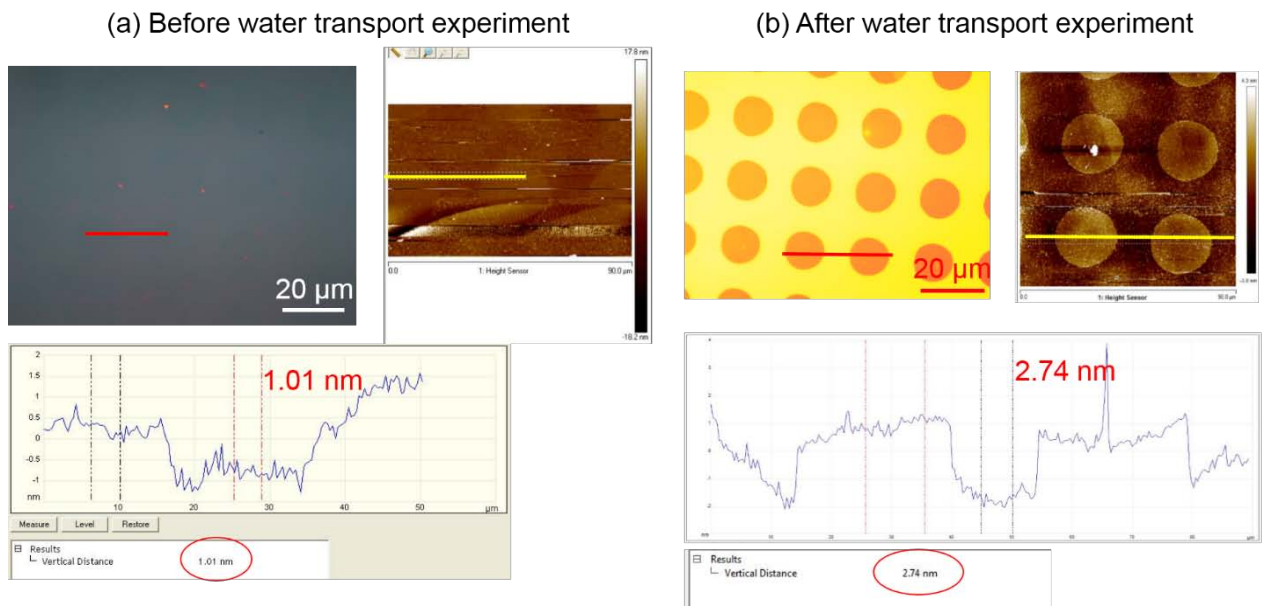

**Figure S11: AFM measurements taken before and after experiments of water transport through polymer nanomembrane. a.** shows the optical image of the polymer NM patterned on a 30nm Ge device which is invisible at the start of the etch. The thickness of the polymer measured is 1.01 nm. **b.** shows optical image of the polymer NM patterns clearly visible at the end of the etching process. The AFM measurement shows a total thickness of 2.74 nm measured across the patterns, which includes the 1 nm thick polymer and the height difference in Ge film due to the differential etch rate under the polymer.

**SVideo 2** shows the complete time course of the etching process. Some snapshots from the video are provided in Fig. S12 below.

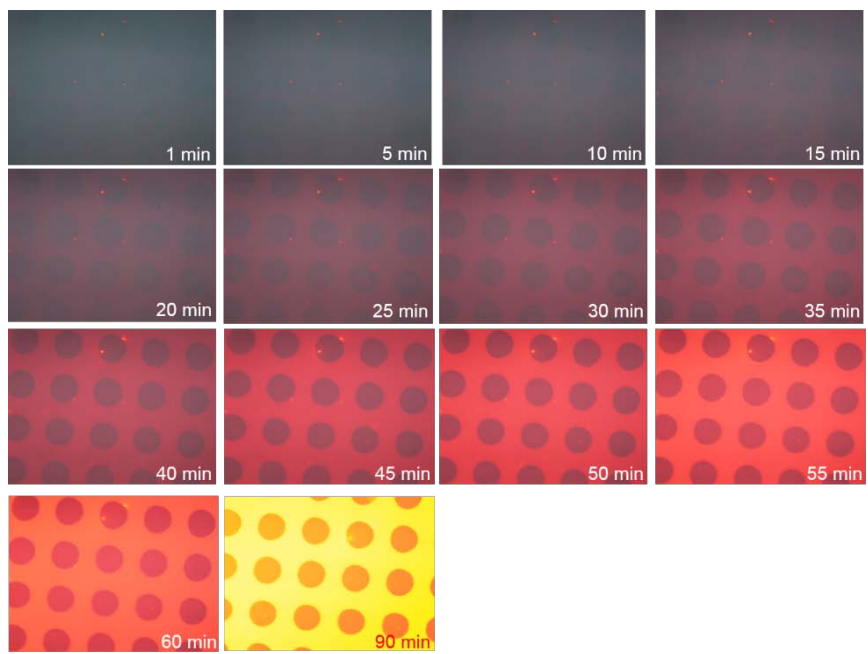

**Figure S12:** Time course of water transport through a single layer polymer of 1 nm thickness. The time at which each image was captured is indicated below in each of the images.

**Section 10: Scanning Electron Microscope (SEM) images of Ag NPs**

For the visualisation of nanoparticles on Ge devices using our technique, we used silver (Ag) nanoplatesof lateral dimension around 40 nm from Nanocomposix Inc. The concentration of Ag NPs used was $21 \times 10^{-14}$ gm/mL ($10^9$ Ag NPs/ mL) in DI water with a pH maintained at 9.3. The SEM image of the Ag NPs dispersed on a Piranha cleaned bare Si wafer (Fig. S13) confirms the size of single NPs around 40 nm. The SEM used was Zeiss Gemini SEM model.

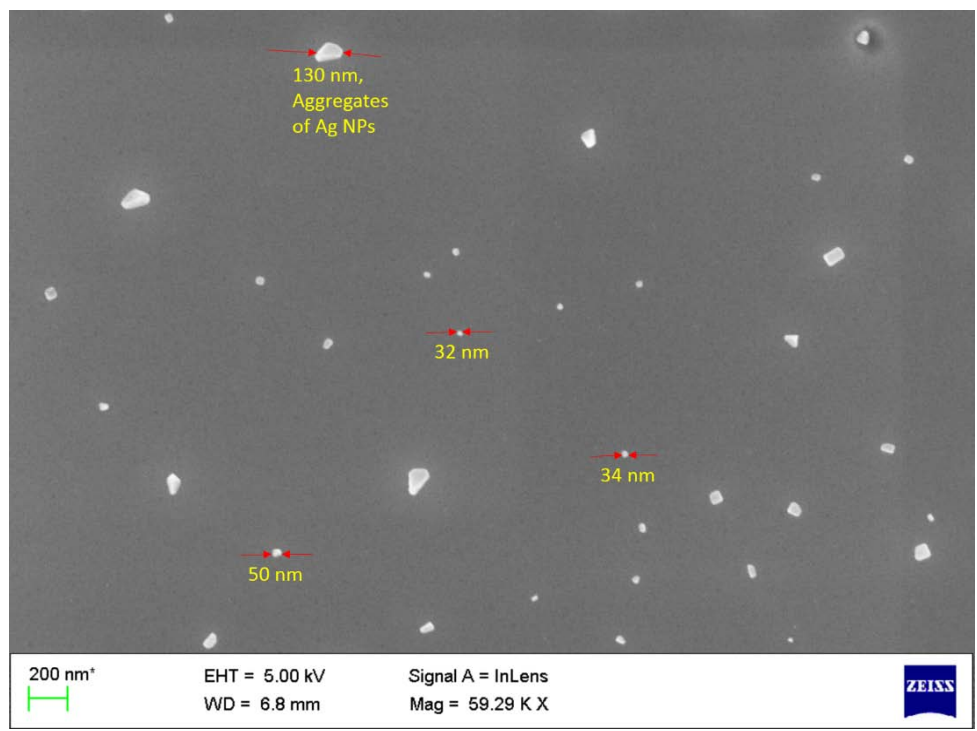

**Figure S13:** Electron Microscope image of Ag NPs drop casted on Silicon.

**Section 11: Visualization of single CNTs using differential etching**

In order to probe the ultimate limit of detection/visualization using this technique, we attempted to visualize Single Walled Carbon Nano Tubes (SW-CNTs). A solution of SW-CNTs with concentration 10ng/ mL containing CNTs with diameter 1-2 nm and length 2-3 microns was used. A 10 µL of this solution was drop casted on the device and left to dry at room temperature so that the SW-CNTs settle and adhere to the Ge device. Water etching experiments carried out on these samples revealed certain features in the Bright-field optical images correlated with Atomic Force Microscopy (AFM) images to be single SW-CNTs (Fig. S14 (a)). Figure S14 (b) and (c) are BF images after image processing where two well separated CNTs are identified while (d) and (e) are the AFM 2D scan and height profile across the CNT marked in (c) and (d). The location of the SW-CNTs are indicated by the red solid lines in Fig S 14 (c) and yellow solid lines in Fig S 14 (d). The optical contrast was quite poor in these experiments. We believe that the poor contrast could be due to movement of CNTs during the rather long etching process. Spatial movement of the CNTs will cause the contrast to get averaged out. Improving the adhesion of the CNTs to the surface using functionalization layer and/or modifying the Ge film to achieve much faster etch rates is likely to improve the optical contrast.

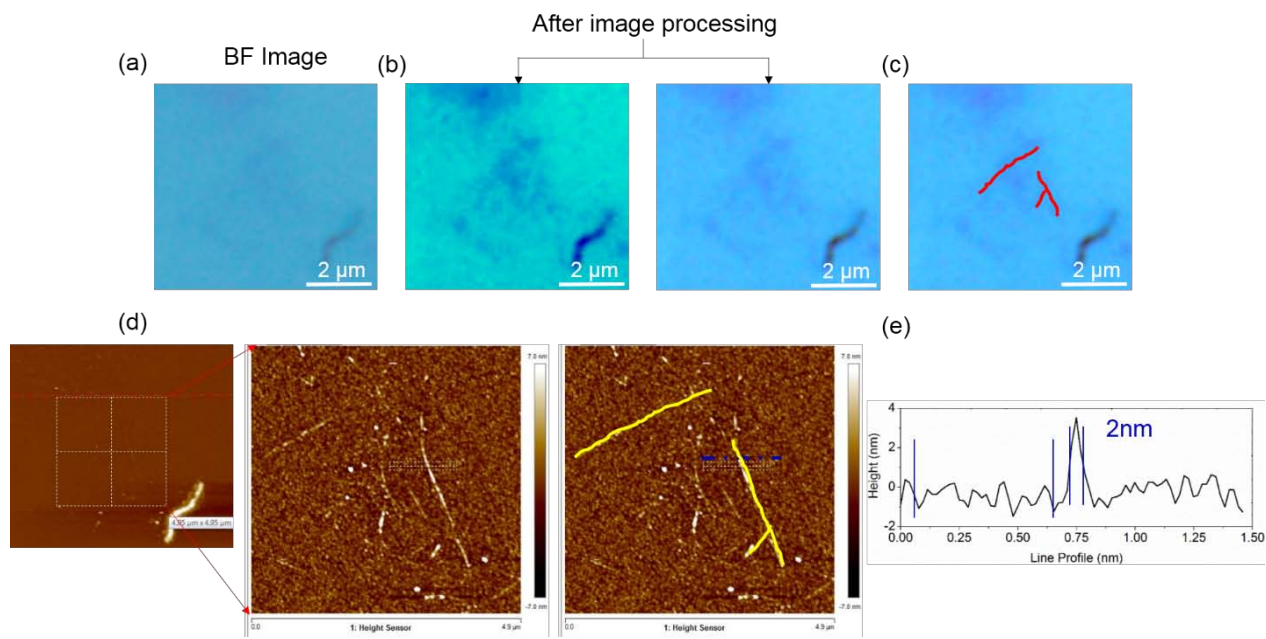

**Figure S14: Visualizing single CNTs. a.** Optical image captured in the BF mode. **b.** and **c.** BF image in **a.** after image processing. The location of the isolated CNTs are indicated by the red solid lines drawn over them. **d.** 2D AFM map of the corresponding location in **a.e.** Height profile indicating the CNT's height measured across the blue dotted line in **d.**

**Section 12: Measuring transport of ions using ion-selective layers**

As mentioned in the discussion section of the main text, one obvious limitation of this technique is that it is currently restricted to probe transport of water alone. However, by using polymer thin films which selectively etch in the presence of specific ions, it will be possible to measure the transport of these ions. As an example, consider the measurement of Na+ ion transport through 2D membranes. In this case, $Ca^{2+}$ conjugated poly-acrylic acid (PAA-$Ca^{2+}$) can act as an ion selective layer as the solubility of PAA-$Ca^{2+}$ increases with increasing $Na^+$ concentration in water [Ref. 28 in the main text]. By depositing an intermediate layer or PAA-$Ca^{2+}$ sandwiched between the nano-membrane and the Ge film as shown in Fig. S15 below, one can probe the transport of $Na^+$ ions through the nanomembrane. For the film structure shown in Fig. S15, consider that the nanomembrane impedes the transport of $Na^+$ ions. In that case the ion selective PAA-$Ca^{2+}$ layer would dissolve at a faster rate outside the area covered by the membrane leading to a differential etch rate and Ge film thickness difference as shown in Fig. S15. The relative color difference between the area covered by the membrane and the background is a measure of $Na^+$ ion transport through the membrane. Quantitative models can then be employed to extract relevant transport parameters.

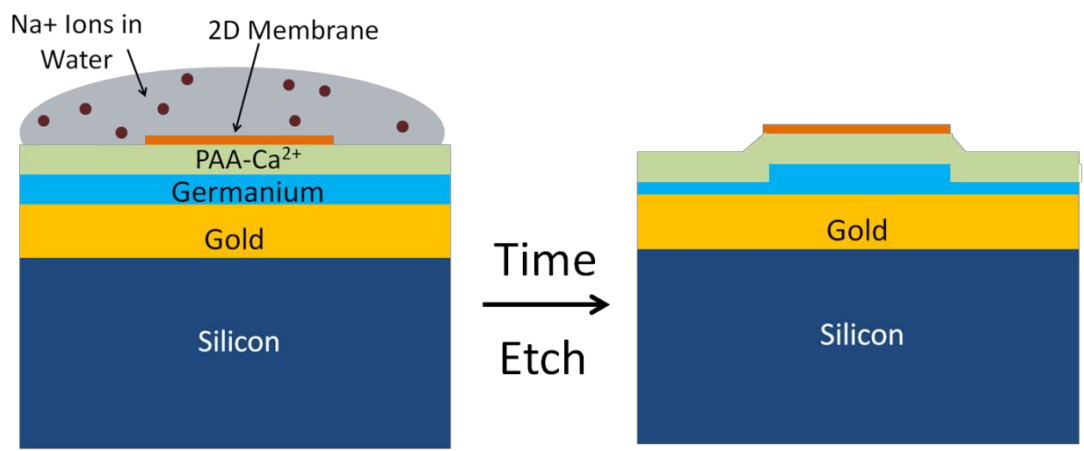

**Fig. S15:** Schematic illustration of measuring Na+ ion transport using an ion selective PAA-$Ca^{2+}$ intermediate layer.